# Shake It or Shrink It: Mass Transport and Kinetics in Surface Bioassays Using Agitation and Microfluidics

*Iago Pereiro[§], Anna Fomitcheva-Khartchenko[§] and Govind V. Kaigala\**

*IBM Research – Europe, Säumerstrasse 4, Rüschlikon, CH-8803, Switzerland*

§ *Both authors contributed equally to this work*

\* *Corresponding author: gov@zurich.ibm.com*

## Abstract

*Surface assays, such as ELISA and immunofluorescence, are nothing short of ubiquitous in biotechnology and medical diagnostics today. The development and optimization of these assays generally focuses on three aspects: immobilization chemistry, ligand-receptor interaction and concentrations of ligands, buffers and sample. A fourth aspect, the transport of the analyte to the surface, is more rarely delved into during assay design and analysis. Improving transport is generally limited to the agitation of reagents, a mode of flow generation inherently difficult to control, often resulting in inconsistent reaction kinetics. However, with assay optimization reaching theoretical limits, the role of transport becomes decisive. This perspective develops an intuitive and practical understanding of transport in conventional agitation systems and in microfluidics, the latter underpinning many new life science technologies. We give rules of thumb to guide the user on system behavior, such as advection regimes and shear stress, and derive estimates for relevant quantities that delimit assay parameters. Illustrative cases with examples of experimental results are used to clarify the role of fundamental concepts such as boundary and depletion layers, mass diffusivity or surface tension.*

## Introduction

Surfaces are ubiquitous to a wide range of popular molecular assays such as ELISA, microarrays[1], immunocytochemistry and immunohistochemistry[2]. In such assays, a key step is the separation of free and bound analytes, for which part of the assay molecules are generally fixed to a solid surface, greatly facilitating the washing and detection steps[3]. The success of this approach has resulted in most immunoassays nowadays relying upon some type of solid phase[4] with a great variety of ligand-receptor couples, labeling formats and step configurations. Additionally, some cell-based assays involving adherent cells are also performed on surfaces, for example shear stress assays to regulate cell morphology and expression[5]. To refer to all such assays performed at the interface of a liquid and a flat solid phase, we use the term "surface assay" here.

The reaction kinetics of molecular surface assays are often significantly influenced by mass transport. Mass transport in a liquid is comprised of both diffusion, driven by concentration gradients, and advection, the transport due to flow bulk motion. The role of mass transport in surface assays is to efficiently bring analytes close to the reactive surface, ensuring that the reaction rate does not decay with time. Because diffusion alone is often insufficient to achieve this, advection is actively generated. The most prevalent advection-generating system are orbital shakers, the standard mixing solution for bioassays, as they are easy to use, widely available, require little supervision and are compatible with standard biological supports, such as microtiter plates[6]. However, shaker-induced advection relies on the sloshing of the liquid free surface. In both our own experience and that of academic and industrial partners we work with, this frequently translates into inconsistent results, non-uniform kinetics and transport mostly limited to the liquid bulk, far from the solid phase. It is therefore not surprising that, from the available protocols for surface assays, it often remains unclear whether shaking is even needed at all. Some reports affirm that the signal improves, while others disregard shaking as 'useless' (to underline this disparity, we have compiled in **Supplementary Table 1** commonly used and proposed conditions).

In this perspective, we aim to clarify when and how advection may be used to improve surface assays from an engineering standpoint. We see this as a rule of thumb guide that illustrates fundamentals and their implementation in the context of two key flow-generating technologies: shakers and microfluidics. The latter field, championed by the chemical world for the last 30 years ago[7], is driving a silent revolution in the life sciences, with numerous applications being reinvented or newly developed at the small scale. These range from the culture of organs-on-a-chip [8] and the separation of biological entities [9] such as blood constituents, to point-of-care devices [10]. Compared to shakers, in microfluidics, liquids are geometrically constrained to the sub-millimeter scale producing a steadier and deterministic advection. Because analytes are brought to the micron-sized fluidic layers adjacent to the solid-liquid interface, this can result in gains in surface kinetics of several orders of magnitude, as compared with no flow conditions[11].

We hope that this perspective will equip chemists and life scientists with a practical understanding of available advection-generating tools to enhance surface assay performance. For this, in **Advection over solid surfaces** we first present the basic underlying principles of hydrodynamics in shaking and microfluidic systems that are relevant for surface assays and directly applicable to e.g. shear stress assays on cells. We particularly focus on how advection, i.e. liquid flows, can be generated in the sub-millimeter vicinity of surfaces in a controlled and predictable way. In **Advection-enhanced surface kinetics** these principles are then used to obtain flow-enhanced kinetics for molecular assays. We focus on operation regimes and intuitive calculations, and both sections include Boxes with illustrative experimental results and calculation examples. Finally, in **Putting flow into effect for surface assays** we discuss and compare the specific use cases and the choice of a system and operating conditions depending on desired results, functionality and availability.

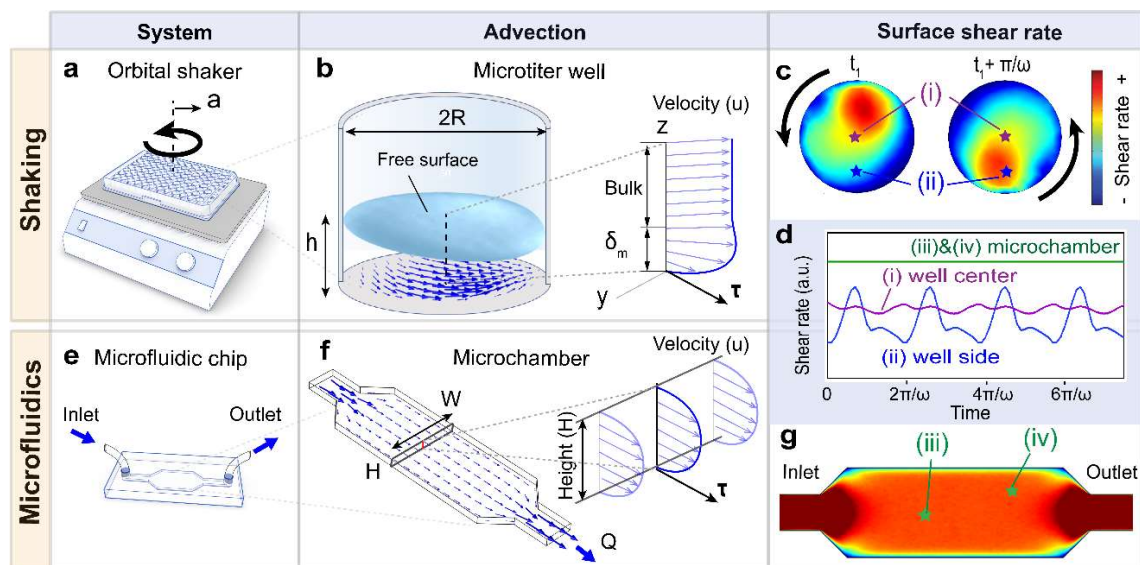

*Figure 1.* *(a) Orbital shaker with a 96 well microtiter plate; (b) example well of a microtiter plate showing a free surface of liquid and advection close to the surface (blue arrows), together with the structure of the velocity profile in the central axis; (c) top view of the well bottom showing simulated surface shear stress for two half cycles: the evolution of shear rate with time at the well center and side positions are shown in (d); (e) microfluidic chip containing a single wide chamber, one inlet and one outlet; (f) flow distribution inside the chamber of a microfluidic chip, with an enlarged view of the Poiseuille flow profile**;** (g) shear rate on chamber bottom, uniform except near the side walls and inlet/outlet apertures, as shown in (d).*

## Advection over solid surfaces

Even when a liquid flows over a solid surface, at the fluid-solid interface the fluid is generally static. This is because in most cases the adhesive forces between the solid and adjacent fluid are greater than the cohesive forces between fluid particles. The result is a gradient of velocities, termed **shear rate** $\dot{\gamma} = \frac{\partial U}{\partial y}$, where $y$ is the direction perpendicular to the surface. The difference in velocity between adjacent liquid layers creates friction through a transfer of momentum, quantified as **shear stress** using units of pressure. For Newtonian fluids, like water and physiological buffers, the shear stress is then simply the shear rate at the interface (y = 0) multiplied by the dynamic viscosity of the liquid (μ):

$$\tau = \mu\,\dot{\gamma} = \mu \frac{\partial U}{\partial y}\bigg|_{y=0} \tag{1}$$

With the shear rate and shear stress we can estimate the flow profile in the vicinity of the interface and the hydrodynamic forces experienced by any object on it, respectively. Therefore, they are critical to understand and describe the surface hydrodynamics of orbital shakers and microfluidic systems, as shown in the next section. Additionally, in this perspective we will use the term advection to refer to the mass transport driven by flows. The other main form of transport in liquids is diffusion, driven by concentration gradients.

## Advection in orbital shakers

In orbital shaking, the well or flask containing the liquid follows a circular trajectory at constant angular velocity ω with a fixed orientation with respect to a static frame of reference. Therefore, an orbital shaker imparts the same in-plane velocity to all points:

$$\vec{u} = a\omega \sin(\omega t)\,\vec{i} + a\omega \cos(\omega)\,\vec{j} \tag{2}$$

where $a$ is the orbital radius. In experimental practice the shaking frequency n = $\omega/2\pi$ is a common parameter (with Hz or rpm as a unit).

Bulk mixing is mostly a result of the movement of the liquid free surface. This liquid-air interface undergoes inclination changes and with enough angular velocity deforms into a surface wave with its leading edge moving along the sidewalls in a sloshing motion (Fig. 1b). Due to its inherent complexity, no model yet exists to capture the complexity of advection in orbitally-shaken containers. It would have to account for the curvature of the free surface, the possible breaking of the surface wave or the presence of crested wave patterns[12], among others. Generally, advection in microtiter wells is approximated by considering two regions: a bulk phase, containing most of the liquid and moving at a constant velocity, and a boundary layer below within which the velocity decreases towards the bottom surface (Fig. 1b). The thickness of this layer is of the same order of magnitude as the viscous penetration depth:

$$\delta_m \approx \sqrt{\frac{\nu}{\omega}} \tag{3}$$

Thus, for a given viscosity, a faster angular velocity $\omega$ leads to a reduction of this flow-dumped area. This will prove to be very relevant for the kinetics described in **Advection-enhanced surface kinetics**. For the magnitude of the shear stress at the well bottom, the following expression, derived from Stokes second problem, is commonly employed, particularly in shear stress studies on adherent cells (see **Box 1**).

$$\tau = a\sqrt{\rho\mu\omega^3} \tag{4}$$

Unfortunately, Eq. 4 does not capture spatial dependence and the estimated shear stress values sometimes differ considerably from experimental observation. As exemplified in Fig. 1c for a microtiter plate well, the orbital movement of the bulk typically translates into a non-uniform and time-dependent advection at the bottom surface. More specifically, the shear rate is maximum where the free surface is shallowest, at the leading edge of the wave[13]. This maximum rotates with $\omega$ (Fig. 1d) so that at any fixed point on the surface the magnitude of the shear rate oscillates with that frequency at different amplitudes (Fig. 1e). In general, shear stress is more constant at the center[13] and decreases in the close vicinity of the sidewalls. Overall, the shear rate is influenced by the orbital speed, well diameter, viscosity, radius of the orbit, fluid height, density and gravity[13].

---

**Box 1: Shaking-induced shear stress on adherent cells**

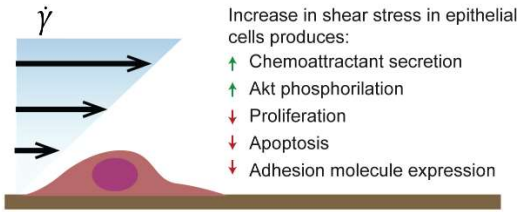

Increase in shear stress in epithelial cells produces:
↑ Chemoattractant secretion
↑ Akt phosphorilation
↓ Proliferation
↓ Apoptosis
↓ Adhesion molecule expression

Advection-driven shear stress on eukaryotic cell cultures can induce morphological changes[14], affect proliferation[15], activate signaling pathways[16] and alter gene or protein expression[14]. In an organism, an example of cells exposed to shear stress are the endothelial cells coating blood vessels. Such cells behave differently if not subjected to the expected shear [17], compromising *in vitro* results. A common tool to simulate shear stress artificially are orbital shakers, which allow parallelization for prolonged experimental times. Unfortunately, the obtained shear stress is often significantly lower in the center than near the side walls[14,16,18–22] leading to a spatial variation in cell morphology [14,22], permeation, proliferation, apoptosis rates or the expression of biomolecules [14–16,21,23]. Although these location-dependent differences might be advantageous in some cases[18], it is important to note that the shear stress calculated with Eq. 3 can in some cases differ from actual values by one order of magnitude in regions far from well center[24] and is most accurate for high frequencies[25].

---

Some theoretical models employing potential flow[24] can more accurately estimate values of surface velocity and free surface deformation [12,26] but so far all available models are limited to certain flow regimes. Consequently, working with dimensionless numbers is generally favored, to more easily link observed advection regimes to experimental conditions. A few well-established dimensionless numbers are the Reynolds number, defining a laminar-turbulent transition, the Froude number, proportional to the steepness of the free surface, and the phase number, which indicates when the liquid is in phase with the shaking movement. Some relevant dimensionless numbers are listed in **Supplementary Note 3**.

For small liquid volumes the surface tension can represent an important resistance to the deformation of the free surface, severely limiting advection. This surface tension needs to be overcome by the labor delivered by the centrifugal force, at a **critical shaking frequency** [27]:

$$n_{crit} = \sqrt{\frac{\sigma R}{4\pi V \rho}} \qquad (5)$$

Where *R* is the well radius, σ the surface tension of the air-liquid interface, and *V* the filling volume. If n < $n_{crit}$ low advection and mixing occurs, whereas mixing increases exponentially above this value [28]. The smaller the shaking diameter, the higher the frequency must be to obtain mixing[29] so that $n_{crit}$ is most relevant for 48, 96 and 384 well plates [29] and becomes less important for 24 well plates and greater [30] [31]. In general, the consideration of parameters such as $n_{crit}$ together with dimensionless numbers such as Reynolds or Froude, is critical to obtain a favorable mixing of the bulk phase [29] and this is key for surface kinetics as shown in **Advection-enhanced surface kinetics**.

**Advection in microfluidic devices**

In contrast to shaking-driven advection, microfluidic flows are simple to describe and control, with well-established analytical models. Because of the dimensional confinement of the liquid in microchannels and microchambers (Fig. 1e), advection in microfluidic systems is mostly restricted to a laminar regime (see Reynold's number $Re_m$ in **Supplementary Note 4**). Thus, fluid particles move in smooth layers, with these layers moving over adjacent layers without mixing, making particle trajectories highly predictable.

Additionally, microfluidic advection over flat substrates is typically created with rectangular-shaped channels or chambers. For these and many other shapes[32], exact analytical solutions to the flow field can be derived from the continuity and momentum equations for incompressible Newtonian fluids. If the channel height $H$ is much shorter than the channel width $W$, the profile of velocities is well defined by a plane Poiseuille flow (Fig. 1f):

$$u_x(z) = \frac{1}{2\mu}\frac{dp}{dx}z(H-z) \qquad (6)$$

With a purely unidirectional flow velocity, maximum in the channel center ($z = H/2$) and with a value of zero at the walls ($z = 0$ and $z = H$). Note that this profile is independent of the transversal direction y and thus valid in the full width of the channel except close to the sidewalls. As seen, the only parameters required to create a fully controlled advection are the channel dimension, the viscosity of the liquid and the pressure gradient along the channel dp/dx. In practice, this pressure gradient is imposed by an external pressure pump or syringe-injection systems.

From Eq. 1 and Eq. 6, and considering the flowrate through the channel section, $Q = W\int_0^h u\,dz$, the shear rate on the substrate is:

$$\dot{\gamma} = \frac{6Q}{WH^2} \qquad (7)$$

Hence, the applied shear rate is experimentally defined with two parameters, the flowrate of the liquid and the vertical dimension of the channel. The latter is generally fixed by the chip geometry while the former can be modified during the experimental run.

This obtained shear stress is fully uniform over the substrate except (i) in close vicinity to the sidewalls, and, (ii) at channel entrance as flow develops into a Pouiseuille flow (Fig. 1g). The size of both regions is typically in the order of a few channel heights and can be analytically calculated if needed (see [32]).Thus, to maximize the shear rate/stress on a substrate the correct geometry can be defined with a few simple rules. In **Box 2**, the shear rates from both shaking and microfluidics are illustrated indirectly through experimentally-obtained surface kinetics. Note that shaking-induced advection results in a non-uniform assay signal in this case. As seen next, this points to a lower shear rate in the center of the wells, in agreement with the lower maximum values of shear seen in Fig. 1c.

In this section we have seen how advection and its derived shear stress is generated on flat surfaces though shaking and microfluidics. This can be directly applied to some assays that rely on shear-induced effects. An example of these are shear stress assays on adherent cells, as shown in **Box 1**. The case of analytical surface assays is treated in the next section, where these same principles are used to understand and estimate the effect of advection in the reaction kinetics.

Box 2: Advection-enhanced surface kinetics in a microtiter plate

Consider a well of a 96-well microtiter plate containing 100 µL of phosphate buffer saline (PBS). For this well size, surface tension effects are dominant at low shaking frequencies (see **Advection in orbital shakers**) and, for an orbital radius a = 1 mm, the critical frequency is $n_{crit}$ = 810 rpm (Eq. 5). As shown in (a) of the figure below, the experimental behavior of the free surface exhibits proportionally small level changes up to 750 rpm. Further, in (b) the well bottom is coated with an IgG molecule and the liquid sample contains a fluorescent anti-IgG antibody and is left to shake at the corresponding frequencies for 5 minutes (see **Supplementary Note 1** for experimental details). The resulting image intensity correlates with the experimental shear rate (see **Surface kinetics in orbital shakers**). We see that increasing frequencies lead to higher intensities but also to a signal gradient between center and well side (~37% variation at 1000 rpm) due to local differences in shear rate (for 1500 rpm transport is so efficient that the signal saturates).

Consider now in (c) the application of a micro-scale laminar advection on the same substrate by introducing a microfluidic system in the well (see **Supplementary Note 1**), creating a space of height H = 30 µm where the same anti-IgG sample is flown. The shear in this case is much more uniform (d) resulting in a significantly flatter intensity profile after 5 minutes at Q = 5 µL/min (intensity variation ~5%). See **Box 3** for details on the kinetics.

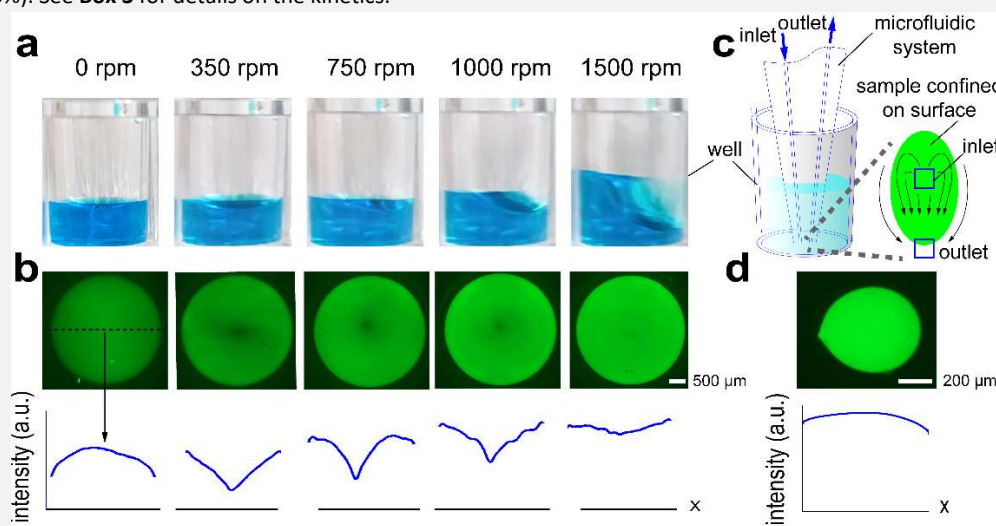

## Advection-enhanced surface kinetics

A key application of advection in chemistry and the life sciences is the kinetic enhancement of chemical reactions on substrates. Applications include sensors and surface assays, in formats such as receptor-ligand and molecular adsorption. In all of these cases the process takes place in two steps: (i) the transport of analytes to the substrate, and (ii) the chemical reaction on the surface. Depending on their rate the overall process can be limited by either one of the two steps. Considering Fick's first law for (i) and first order reaction kinetics for (ii), the ratio of the reaction rate to the mass transport rate is the Damköhler number (**Supplementary Note 5**):

$$Da = \frac{k_{on} b_m \delta_c}{D} \qquad (8)$$

where $k_{on}$ is the analyte-receptor forward kinetics, $b_m$ the density of surface receptors and D the diffusivity constant of the transported analyte. If Da >> 1, kinetics are transport-limited, so that the reaction consumes analyte much faster than transport can provide it, forming a layer of low analyte concentration near the surface, termed depletion zone $\delta_c$. If Da << 1 kinetics are reaction limited and mass transport is not relevant. In static systems where transport is limited to diffusion, the depth of the depletion layer is given by Einstein's equation for diffusion $\delta_c = \sqrt{2Dt}$ and Da thus increases with time. Unless the initial Da << 1, the chemical reaction rate progressively decreases because it becomes increasingly transport-limited. This is the case of most ligand-receptor systems in the life sciences that use relatively large molecules with D = $1.10^{-(10-13)}$ m$^2$/s including albumins, enzymes, antibodies or DNA [33,34]. The objective of advection, both with shaking and microfluidics, is to reduce Da to enhance transport by **reducing the depletion zone** (Fig. 2a), thus unrestraining the chemical reaction (**Supplementary Figure 1**).

### 3.1 Surface kinetics in orbital shakers

In shakers, a fixed liquid volume stays on top of the substrate and mixing keeps the depletion layer from growing into the bulk. We saw in **Advection in orbital shakers** that advection in shaken wells can often be approximated by a boundary layer of thickness $\delta_m$ over a bulk liquid layer. Helpful here is the dimensionless Schmidt number, which compares momentum diffusivity to mass diffusivity Sc = v/D. It relates the relative thickness of the boundary layer to the diffusion-defined depletion layer. In liquids the Schmidt number is generally high (Sc >> 1) and in the laminar regime the ratio between layers is in the order of: [35,36]

$$\frac{\delta_m}{\delta_c} \approx Sc^{1/3} \tag{9}$$

From this, we can estimate the depletion zone as a fraction of the height of the boundary layer, which itself is a function of the rotation speed and liquid viscosity (Eq. 4). With $\delta_c$, we can estimate the potential gains obtained by shaking our surface assays (see **Box 3**). Caution should however be taken because of the complexity of the system and the approximate nature of value of $\delta_m$ (see **Advection in orbital shakers**). This is particularly visible in Box 2, where experimental results are given for a shaked assay in which a fluorescently-labeled antibody binds to the surface of a microtiter well. The lower signal value in the well center under several experimental conditions points to a non-uniform boundary layer, the values of $\delta_c$ being smaller near the walls. Additionally, not all shaking regimes result in the formation of surface boundary layers. For this, **Supplementary Note 3** examines advection regimes that can support good bulk mixing and thereby the formation of boundary layers.

### 3.2. Surface kinetics in microfluidics

In microfluidics, in the presence of a flow, new sample is continuously fed over the substrate. The Peclet number, relating transport by diffusion to convection, is a useful dimensionless number to indicate at which rate this liquid needs to be provided:

$$Pe = \frac{\tau_d}{\tau_a} = \frac{2Q}{WD} \tag{10}$$

If the time for diffusion to cross the channel from top to bottom $\tau_d$ is shorter than the time for advection $\tau_a$ to move that same distance, i.e. *Pe << 1*, advection may be too low to feed the surface chemical reaction, consuming the analyte at the entrance[37]. If the user wants the binding to occur uniformly across the entire surface of length L, the flowrate Q should be high enough to result in *Pe >> 1*. In this situation, the size of the depletion layer can similarly be calculated as the distance from the substrate at which a particle has the same probability of reaching the substrate by diffusion as of leaving the reaction area by advection[38]:

$$\delta_c = \left(\frac{2DL}{\dot{\gamma}}\right)^{1/3} \tag{11}$$

With the shear stress $\dot{\gamma}$ given by Eq. 7. As expected, higher shear rates result in smaller depletion zones, leading to a shift of the Damköhler number towards reaction-limited conditions (Eq. 8). Thus, when flows are present the size of the depletion layer eventually stabilizes, and thereafter the transport rate does not vary with time until saturation[38]. For the flow conditions considered in **Advection in microfluidic devices** this depletion layer is also relatively uniform across the binding surface, as illustrated in the example assay of Box 2. Mass transfer through the depletion layer is inversely proportional to its size and thus the main gain of a reduced $\delta_c$ is the shortened time to result of the assays (see **Supplementary Note 5** for calculations of assay times). Further, knowing how to estimate $\delta_c$ we can use it to evaluate the expected performance of advection-generating systems and experimental parameters. This facilitates the user the task of system choice (see **Putting flow into effect for surface assays**). For example, in **Box 3** the kinetics of an antibody-antigen ELISA are compared using an orbital shaker and a microfluidic system. **Supplementary Figure 2** further examines an equivalent case where the analyte constitutes of slower diffusive gold nanoparticles, illustrating how at the sub-micrometer scale lower diffusivities D benefit from increased advection. For larger particles, more prone to sedimentation, the situation changes significantly, as shown next.

### Box 3: Antigen-antibody binding kinetics with orbital shaking and microfluidics

Consider the bottom of a standard 96 well plate with a radius *R = 3.3 mm*. This substrate is covered by IgGs with an approximate density $b_m$ = *18000 sites/µm²* and filled with 100 µL of a solution of anti-IgG antibodies with the affinity $k_{on}$ = 1000 m³/mol.s, $k_{off}$ = 1 $10^{-3}$ s$^{-1}$, a diffusion coefficient $D_{IgG}$ = 6.5 $10^{-11}$ m²/s and a concentration $c_0$ = 1.3 $10^{-8}$ mol/L.

If left static, the depth of the depletion layer increases with time (see **Surface kinetics in orbital shakers**) so that at the characteristic time scale τ of the reaction (see **Supplementary Note 5**) it reaches a size of 96 µm, at which point the Damköhler number (Eq. 8) will be Da = 44. Because Da >> 1 the system is transport-limited and thus advection could help improve kinetics. The figure below shows real experimental results of such a system (materials & methods in **Supplementary Note 1**). If using a shaker of an orbital radius a = 1 mm, the critical frequency $n_{crit}$ ≈ 810 rpm (Eq. 5) is in the upper range of most commercial shakers. Thus, at a much lower 350 rpm, as shown in the figure, there is no significant improvement of kinetics compared with static conditions. At 750 rpm the system already becomes dynamic and we can estimate a boundary layer $\delta_m$ ≈ 113 µm (Eq. 3) and a depletion layer $\delta_c$ ≈ 4.5 µm (Eq. 9). Now Da = 2, a clear improvement of the system kinetics.

Consider now the application of the same sample with microfluidics at H = 30 µm, L = 500 µm and W = 300 µm, (see Box 1). We consider two distinct flowrates, 5 and 100 µL/min. The Peclet number is Pe >> 1 in both cases, enough to compensate for the analyte consumed by diffusion. The depletion layers are $\delta_c$ = 3.3 µm and $\delta_c$ = 1.2 µm, respectively (Eq. 11), corresponding with the observed binding kinetics: at 5 µL/min the kinetics are similar to the ones obtained with the shaker at 750 rpm, improving for 100 µL/min, where Da = 0.5, which is close to the ideal reaction-limited case.

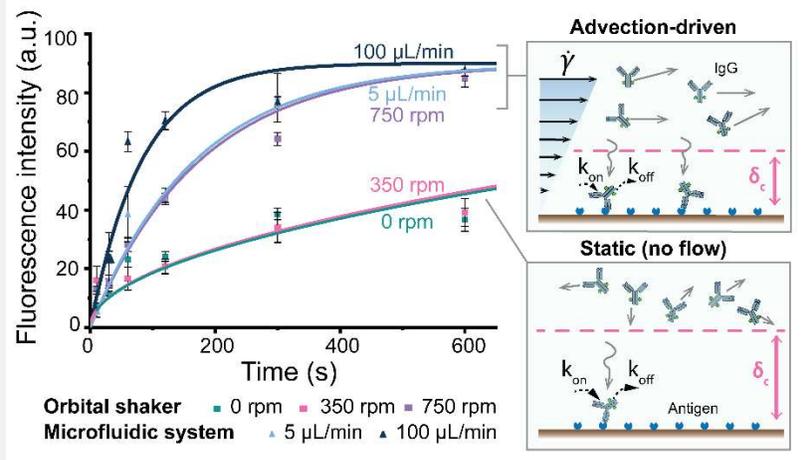

### 3.3. The case of particles larger than molecular size

Particles with a diameter higher than some hundred nanometers are common in surface assays as part of the detection system (e.g. nanoparticles), binding mechanism (e.g. magnetic beads) or the sample itself (e.g. cells). Compared to molecular-sized species, the transport mechanism of these larger particles differs substantially: they present very low diffusivities and, in the absence of advection, they sediment at a settling velocity given by Stoke's law:

$$v_s = \frac{2}{9}\frac{(\rho_p - \rho_f)}{\mu} g R^2 \qquad (12)$$

where $\rho_p$ and $\rho_f$ are the densities of particle and fluid respectively and $g$ is the gravitational field strength. Because the radius is squared here, the effect is almost negligible at the small scale but quickly becomes relevant as size increases if $\rho_p > \rho_f$. Such a velocity will lead to a total mass transfer $j_s = c_0 v_s$ from the bulk to the surface. We can compare this to the transport rate through the depletion layer (**Supplementary Note 5**) resulting in a ratio $\varepsilon = v_s \delta_c / D$, valid for $Da \gg 1$. Thus, for example, for a depletion layer $\delta_c = 100$ *nm* the flux due to sedimentation is negligible for 10 nm gold nanoparticles ($\varepsilon \approx 10^{-6}$) but very relevant for 1 μm particles ($\varepsilon \approx 1$) and predominant for 10 μm particles ($\varepsilon \approx 10^3$).

If advection is present, particles are dragged along by the flow, reducing sedimentation. They can only adhere to a surface if the distance between the two is smaller than the ligand-receptor length, with their binding kinetics depending on particle translational and rotational velocities[39]. In practice, for particles with ε values close to or higher than 1 the attachment rate tends to decrease with increasing advection[40] and therefore this is not beneficial in general. An example of this are red blood cells, as illustrated in **Box 4**. However, some advection is often still used to increase throughput, such as for cell sorting[41], and/or reduce non-specific binding and perform washing steps. For this, it is possible to define an advection regime below a critical shear stress $\tau_c$ in which particles of interest can remain bound in the presence of flows[42]. **Supplementary Note 6** explains how $\tau_c$ is calculated while **Box 4** illustrates the case for red blood cells.

### Box 4: Surface adhesion of cells

Consider the case of red blood cells (RBCs) with a density $\rho_p = 1.11$ g/mL with membrane-bound anti-IgGs. IgG affinity, analyte concentration and surface properties are assumed to be the same as in Box 3. RBCs are non-spherical, but in a first approximation we assume an effective radius $R = 3.3$ $\mu m$[43] and thus use the Stokes-Einstein equation (**Supplementary Note 5**) to calculate their diffusion coefficient, $D = 7.7 \cdot 10^{-14}$. The height of the liquid for a sample volume of 100 µL is 3 mm, thus requiring a time t ≈ 16 min to settle all cells at $v_s \approx 3 \cdot 10^{-6}$ m/s (Eq. 12). The experimental results below show a cell adhesion trend that tends towards saturation around that time. In contrast, when shaking at 750 rpm convection disrupts the settling flux resulting in a much-reduced adhesion rate. This was expected: at this frequency, the theoretical depletion layer would be very small $\delta_c \approx 0.5$ $\mu m$ but $\varepsilon = 19 >> 1$ shows that settling is the dominant transport mode.

Once attached, how much advection can we use to detach unbound cells and/or replace the liquid? Estimating the density of antibodies on the cell membrane as $N_L = 100$ sites/um$^2$ and $K = 1.4 \cdot 10^{-10}$ cm$^2$, we can again consider RBCs as being spherical to calculate a critical shear stress $\tau_c \approx 8.2$ Pa (**Supplementary Note 6**, see [44] for advection-driven forces on oblate particles). In principle this would mean that, for a = 1 mm, the shaker could run up to n ≈ 4000 rpm without cell detachment (Eq. 4), although in practice the liquid would probably already leave the well center in favor of the walls at lower frequencies. In the case of microfluidics, for the same configuration as in Box 3, reagents could flow up to a flowrate limit of Q ≈ 25 µL/min (Eq. 7).

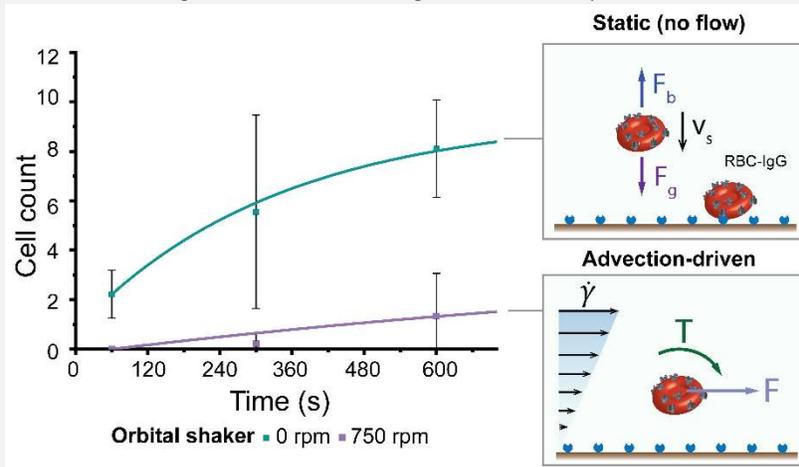

## 4.0 Putting flow into effect for surface assays

Advection is most beneficial when kinetics are otherwise transport-limited (Da > 1) and sedimentation is not the main mode of mass transfer to the target surface ($\varepsilon$ < 1). To choose a system to enforce advection, Fig. 3 aims to guide the user on the main characteristics of orbital shakers and microfluidic systems, compared in a spider plot.

Orbital shakers are strongest in terms of throughput and ease-of-use. A key attribute of orbital shakers is their compatibility with flasks, Petri dishes and microtiter plates, the standard supports for enzymatic assays, toxicity tests, drug screening or microorganism culture[45]. This allows easy multiplexing and parallelization, although reagent volumes cannot be reduced beyond dozens to hundreds of microliters. As shown in Box 1, the signal obtained often exhibits a radial symmetry. To overcome this, arrays of spots in a circular layout may provide a more reliable signal quantitation for surface assays[35]. As seen in **Advection over solid surfaces** however, shaker-driven advection is inherently complex and difficult to generalize to other scales. For example, Alpresa *et al.* identified no less than 12 different flow regimes restricted to shallow liquids (Γ<< 1, see SI) in 6-well microtiter plates. To set operating conditions some relevant dimensional numbers can provide a first guidance, and this is examined in **Supplementary Note 3**. However, we see this as being of limited use, as the user still needs to rely on the literature of similar geometries and operating conditions and/or perform at least preliminary tests to verify the expected results.

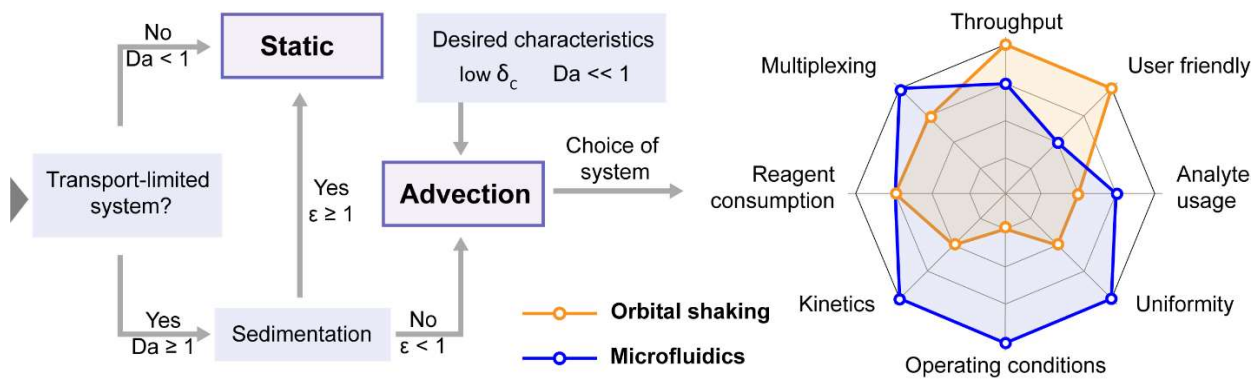

*Figure 2. Flowchart showing the conditions under which advection is beneficial for improving surface kinetics and a comparison of orbital shaking and microfluidics based on technical parameters. Strengths are relative to the best performing system for each parameter, with the higher value being located at the periphery of the plot.*

In contrast, microfluidics offers high predictability: laminar flows are almost inherent to the microscale and the user only needs to ensure that that analyte reaches the substrate (Pe >> 1). If possible, wide channel sections are used (W>>H) to minimize side wall effects and enhance uniformity. Under these conditions, only one advection regime is possible, the depletion layer $\delta_c$ can be minimized with only 2 parameters (H and Q, Eq. 11) and analyte transfer to the surface is well approximated (**Supplementary Note 5**). Additionally, the miniaturization and design flexibility of microfluidic systems can offer unmatched multiplexing, with chips that can include thousands of integrated experimental conditions[46]. Nevertheless, the user should be aware of some caveats of the microfluidic approach, specially a high analyte usage when using strong flows and possible flow-disrupting elements in closed channels, such as bubbles and clogging. In our experience, both can be minimized by implementing appropriate technical solutions, such as recirculation approaches[11] for the former and geometrical features and temperature/pression conditions for the later[47]. In **Box 5**, an example application for blood grouping is illustrated where both shaking and microfluidics are shown to offer advantages in different scenarios.

> Box 5: Example application – Solid phase assay for blood grouping
>
> Knowing a patient's blood group is an essential requirement in blood banking and prior to transfusion. For this, both the antigens present in red blood cells and antibodies in the plasma are tested, referred to, respectively, as antigen typing and antibody screening. We consider the possibility of applying advection to these tests in a solid phase assay format in three scenarios here:
>
> (i) In antigen typing, micron-sized RBCs bind to an antibody-covered surface. In Box 4, this kind of scenario was already determined ] to not benefit from advection in general (instead, centrifugation is often used to enhance sedimentation).
>
> (ii) A common antibody screening assay consists in testing the presence of RhD blood group antibodies. For example, a woman who is RhD- might develop antibodies against an RhD+ fetus, thus endangering the pregnancy. This type of assay is very similar to the one described in Box 3, with the difference that antibodies bind to antigens present on RBCs immobilized on a surface. These antibodies are generally of high affinity and the antigen is highly expressed on the RBC membrane (high $k_{on}$ and $b_m$). This tends to create transport-limited situations in assays with no flows (high Da, Eq. 8). Here, orbitally-shaked multi-well plates could provide the high throughput requirements, with the enhanced kinetics leading to fast time to results.
>
> (iii) Patients with a history of blood transfusion require testing for immunity developed as a result of previous transfusions. These are antibodies against antigens that generally present a lower density in the RBC membranes, such as Kell or Lewis. These IgGs generally present lower affinity and concentration. However, because sensitivity here is critical, advection could provide the critical gain to reach the more challenging limit of detection. In this scenario of lower throughput and need for high kinetic gains, disposable microfluidic devices could be a fitting solution. Such devices could additionally provide multiplexing capabilities, such as with arrays of spots, to test many rare blood groups from a single sample.

Beyond kinetics enhancement, pure shear rate control, e.g. for cell culture, follows similar principles. In this case, the inherent oscillatory nature of the shear stress in shakers is sometimes taken as a model for advection conditions experienced by cells coating blood or lymphatic vessels. However, this shear behavior is distributed non-homogeneously, presents meager predictability and the periodic oscillation can considerably differ from simple sinusoids (as exemplified in Fig. 1d). We think microfluidics can provide a distinct advantage here, as most standard pumps can generate periodic shear stress with minimal software control in all kinds of oscillatory patterns.

Based on this, we believe that shakers are and will continue to be useful tools for surface assays, due to their ease of use and compatibility with existing instrumentation. We see them as an ideal tool for the quick testing of experimental conditions, and in general to provide information in a qualitative or semi-quantitative nature. However, their inherent complex advection and low predictability will likely make them unsuitable for the next generation of high-performance applications. We thus believe that the ongoing efforts to improve the accessibility of microfluidics to the non-specialist could provide the necessary lever to obtain the next levels of accuracy and quantitation in diagnostics, pharmacology, biophysical studies and life sciences.

In this perspective we have attempted to treat the most common parameter regimes in orbital shaking and microfluidics when applied to surface assays. Both systems are extremely rich and many relevant aspects cannot be covered in such a short perspective. However, we hope that this work will serve as a resource on relevant mass transport concepts, often hidden to the non-specialist by mathematical complexity. We believe the simplified overview provided here is most useful to intuitively evaluate systems, regimes and experimental parameters, providing guidance towards more efficient assay designs and implementations.

**Supporting information**

Supplementary notes with further explanations, mathematical expressions and tables (PDF)

**Author Contributions**

A.F. and I.P. designed and performed the experiments, I.P. performed simulations, A.F. and I.P. analyzed the data, A.F., I.P. and G.K. wrote the paper

**Acknowledgements**

We thank Prof. M. Bercovici (Technion, Israel), Dr. S. Descroix (Institut Curie, France) and Dr. E. Delamarche for detailed discussions. We acknowledge funding by ERC-PoC CellProbe (842790). We thank Dr. H. Riel for continuous support.

**Competing Interests**

The authors declare that they have no conflict of interest.

**Table of contents graphic**

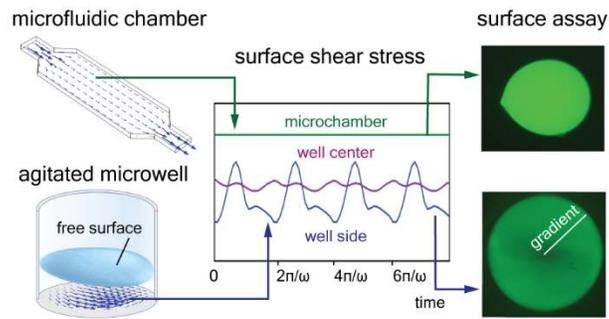